\documentclass[twocolumn,prl,floatfix,superscriptaddress]{revtex4}
\usepackage{graphicx, subfigure, amsfonts,amssymb,amsmath, array, gensymb,fontenc,color, lipsum}
\usepackage{tikz}

\begin{document}

\title{{Even-odd conductance effect in graphene nanoribbons induced by edge functionalization with aromatic molecules: Basis for novel chemosensors}}

\author{Kristi\=ans \v Cer\c nevi\v cs}
\email{kristians.cernevics@epfl.ch}
\author{Michele Pizzochero}
\author{Oleg V.\ Yazyev}

\affiliation{Institute of Physics, Ecole Polytechnique F\'ed\'erale de Lausanne (EPFL), CH-1015 Lausanne, Switzerland}
\affiliation{National Centre for Computational Design and Discovery of Novel Materials (MARVEL),  Ecole Polytechnique F\'ed\'erale de Lausanne (EPFL), CH-1015 Lausanne, Switzerland }

\date{\today}

\begin{abstract}  
We theoretically investigate the electron transport in armchair and zigzag graphene nanoribbons (GNRs) chemically functionalized with $p$-polyphenyl and polyacene groups of increasing length. Our nearest-neighbor tight-binding calculations indicate that, depending on whether the number of aromatic rings in the functional group is even or odd, the resulting conductance at energies matching the energy levels of the corresponding isolated molecule are either unaffected or reduced by exactly one quantum as compared to the pristine GNR, respectively. Such an even-odd effect is shown to originate from a subtle interplay between the electronic states of the guest molecule that are spatially localized on the binding sites and those of the host nanoribbon. We next generalize our findings by employing more accurate tight-binding Hamiltonians along with density-functional theory calculations, and critically discuss the robustness of the observed physical effects against the level of theory adopted. Our work offers a comprehensive understanding of the influence of aromatic molecules bound to the edge of graphene nanoribbons on their electronic transport properties, an issue which is instrumental to the prospective realization of graphene-based chemosensors.
\end{abstract}

\maketitle

\section{Introduction}
\label{intro}

In pursuit of opening a band-gap in otherwise semimetallic graphene for achieving switching capabilities in the resulting logic devices \cite{Novo04,Neto09}, the confinement of charge carriers in one dimension -- as accomplished through the realization of graphene nanoribbons (GNRs) \cite{Son06a} -- was established as the leading approach. Initially produced by means of lithography \cite{Han07}, unzipping of carbon nanotubes \cite{Kosy09} or appropriate cutting of graphene sheets \cite{Ci08}, reliable manufacturing methods  along with practical applications of GNRs remained severely hampered by the inability to reach a full control over their atomic structure. In this vein, substantial progress has been made over the last decade \cite{Cai10a}, when surface-assisted polymerization of carefully chosen precursor molecules has demonstrated a superior route to fabricate atomically precise GNRs with target widths \cite{Chen13,Kimo15}, edge geometries \cite{Vo14,Liu15a}, and dopants incorporation \cite{Clok15,Kawa15}. Additionally, owing to graphene's high transparency \cite{Nair08}, structural flexibility \cite{Lee18}, and excellent electrical conductivity \cite{Novo04}, GNRs have been recognized as potential building-blocks for a wide variety of prospective technologies, including components for carbon-based nano- \cite{Ares07,Kang13} and opto-electronics \cite{Chon18a,Suzu18a}, catalysts for hydrogen-fuel cells \cite{Zhen14a}, drug delivery \cite{Mous19a}, as well as gas-sensing devices \cite{Mehd17a}.  

Chemical functionalization with aromatic molecules has been theoretically shown to largely influence the electronic structure of both armchair- and zigzag-edge graphene nanoribbons (AGNRs and ZGNRs, respectively) \cite{Rosa08a,Rosa08b}, such that, upon binding, each functional group leaves a unique "fingerprint" reflecting its energy levels. This finding makes GNRs appealing platforms for the realization of novel chemosensors. 
On the basis of first nearest-neighbor (1NN) tight-binding (TB) Hamiltonians described by the hopping integral $t_1$ between the $p_z$ orbitals, customarily taken to be $t_1=2.70$ eV, it has been predicted that the formation of a chemical bond between GNRs and either $p$-polyphenyl or polyacene groups may affect the conductance spectra at all values of energy matching those of the isolated molecule. Surprisingly, a phenomenon emerges in both AGNR and ZGNR, where the conductance at energy $E=\pm 2.70$ eV is unaltered when the guest molecule features an even number of rings, whereas for an odd number of rings a Fano anti-resonance (stemming from the interference between the continuum of states of the hosting GNRs and the states localized on the guest molecule \cite{Miro10}) occurs and removes one conductance quantum \cite{Rosa08a}. Despite its potential in designing GNR-based sensors for aromatic molecules, such an intriguing effect remains very poorly understood to date.

A previous work \cite{Rosa08a} justified this even-odd effect by suggesting that an isolated molecule featuring an odd (even) number of rings would (not) exhibit such an electronic state located at $E=\pm 2.70$ eV, in close analogy with chains hosting even or odd number of quantum dots \cite{Orle03}. Yet, we remark that polyacenes or $p$-polyphenyls, irrespective of the number of rings, \emph{do present} an energy level at $E=\pm t_1 = \pm 2.70$ eV in their spectra at the 1NN-TB level of theory, thus proving the previous argument false.  Hence, this consideration highlights that the conductance profile of such edge-functionalized GNRs depends only \emph{partially} on the energy levels of the isolated functional group, signaling that more complex effects ensuing from the interaction of the electronic structure of the hosting nanoribbon with that of the guest molecule are operative and have to be clarified. In addition, it remains to be ascertained whether such even-odd phenomenon is actually physical or simply emerges as an artifact of the (possibly oversimplified) 1NN-TB model. 

In this paper, we investigate in detail, at the TB and first principles levels of theory, the effect of $p$-polyphenylene and polyacene molecules of varying length covalently bound to the edge of AGNRs and ZGNRs on the electron transport by means of non-equilibrium Green's function calculations. We first unravel the origin of the even-odd conductance effect upon binding of guest molecules and rationalize the otherwise unclear loss of exactly one conductance quantum in terms of the spatial distribution of the incoming wave function. Next, we extend our findings to higher levels of theory in order to critically examine the robustness of the observed effects with respect to the adopted theoretical models. Overall, our work establishes a comprehensive picture of the impact of the aromatic functional groups on the electron transport properties of graphene nanoribbons, hence contributing to lay the conceptual foundations underlying the practical realization of GNR-based chemosensors.

\section{Methodology}
\label{method}
We obtain Hamiltonians at both tight-binding (TB) and density-functional theory (DFT) levels. Our TB Hamiltonians include one $p_{z}$ orbital per atom and take the general form
\begin{equation}
H=\sum_{i}\epsilon_{i}c_{i}^{\dagger}c_{i}-t_{i,j}\sum_{i,j}(c_{i}^{\dagger}c_{i}+H.c),
\label{eq:hamiltonian}
\end{equation} 
where $\epsilon_{i}$ is the on-site energy at lattice position $i$, $t_{i,j}$ is the hopping integral between the nearest neighbors ($i, j$), and $c_{i}^{\dagger}(c_{i})$ creates (annihilates) an electron at lattice site $i$. We adopt two TB models, both setting on-site energies $\epsilon=0$ eV and including either first nearest-neighbor hopping solely ($t_{1}=2.70$ eV) or up to the third nearest-neighbor hopping ($t_{1}=2.70$ eV, $t_{2}=0.20$ eV, and $t_{3}=0.18$ eV). This latter corresponds to the model proposed and benchmarked by Hancock and coworkers \cite{Hanc10}. All our TB calculations are performed using the \textsc{kwant} package \cite{Grot14}.

Kohn-Sham DFT calculations are performed under the generalized gradient approximation to the exchange and correlation functional devised by Perdew, Burke, and Ernzerhof \cite{PBE}. Core electrons are described by separable norm-conserving pseudopotentials \cite{PSEUDO} whereas single-particle wavefuctions of valence electrons are expanded in a linear combination of atomic orbitals of double-$\zeta$ polarization (DZP) quality. Real space integrations are performed with a 400 Ry mesh cutoff while the Brillouin zone is sampled with the equivalent of 21$\times$1$\times$1 $k$-mesh per unit cell in all cases but transport calculations, for which it is increased to  400$\times$1$\times$1. We optimize the atomic coordinates until the residual force acting on each atom converges to 0.02 eV/{\AA}.  We introduce guest molecules of increasing number of rings in in a 7$\times$1$\times$1 and 14$\times$1$\times$1 supercells of AGNRs and ZGNRs (yielding similar supercell lengths of 30.19 {\AA} and  34.58 {\AA}) containing 126 and 140 atoms, respectively. Replicas along non-periodic directions are separated by a vacuum region larger than 10 {\AA}.

With the tight-binding and Kohn-Sham Hamiltonians $H$ at hand, transport properties are next calculated using the non-equilibrium Green's function (NEGF) formalism
\begin{equation}
\tilde{G}(E)=\left((E+i\eta) I-H-\Sigma_{L}(E)-\Sigma_{R}(E)\right)^{-1},
\label{eq:green}
\end{equation} 

\noindent where $\tilde{G}$ is Green's function, $\eta$ adds an infinitesimally small imaginary part to the energy $E$, $I$ is the identity matrix and $\Sigma_{L(R)}$ is the self-energy representing the semi-infinite left (right) lead. The self-energies are obtained self-consistently

\begin{equation}
\Sigma_{L(R)}(E)=H_{1}^{\dagger}(EI-H_{0}-\Sigma_{L(R)}(E))^{-1}H_{1},
\label{eq:dyson}
\end{equation}

\noindent where $H_{0}$ is the Hamiltonian of the unit cell in the lead and $H_{1}$ is the coupling between the unit cells. The broadening function $\Gamma_{L(R)}$ due to the leads is calculated from the self-energies

\begin{equation}
\Gamma_{L(R)}(E)=i[\Sigma_{L(R)}(E)-\Sigma_{L(R)}(E)^{\dagger}].
\label{eq:gamma}
\end{equation}
\par
\noindent The transmission is obtained by taking the trace of the following matrix

\begin{equation}
T(E)=Tr[\Gamma_{L}\tilde{G}\Gamma_{R}\tilde{G}^{\dagger}].
\end{equation}

\noindent Finally, we use the transmission coefficient  $T(E)$ given above to express the conductivity $G(E)$ in terms of the conductance quantum $G_{0}$ within the Landauer formula

\begin{equation}
G(E)=G_{0}T(E)=\dfrac{2e^{2}}{h}T(E).
\end{equation}
Our quantum electronic transport calculations are performed with the help of \textsc{kwant} \cite{Grot14} and \textsc{transiesta} \cite{TRANSIESTA}.

\section{Results and Discussion}
\label{results}

As compared to their wider counterparts, narrow GNRs are more suitable systems to explore the conductance spectra upon functionalization, as less energy bands are present around $E= \pm2.70$ eV. Hence, in the following we restrict our investigation to 4-ZGNR and 7-AGNR without loss of generality, as the Fano anti-resonances due to the binding molecule occurs irrespective of the number of carbon atoms $N$ across the GNR \cite{Rosa08a}.

\begin{figure*}[t]
	\includegraphics[width=2\columnwidth]{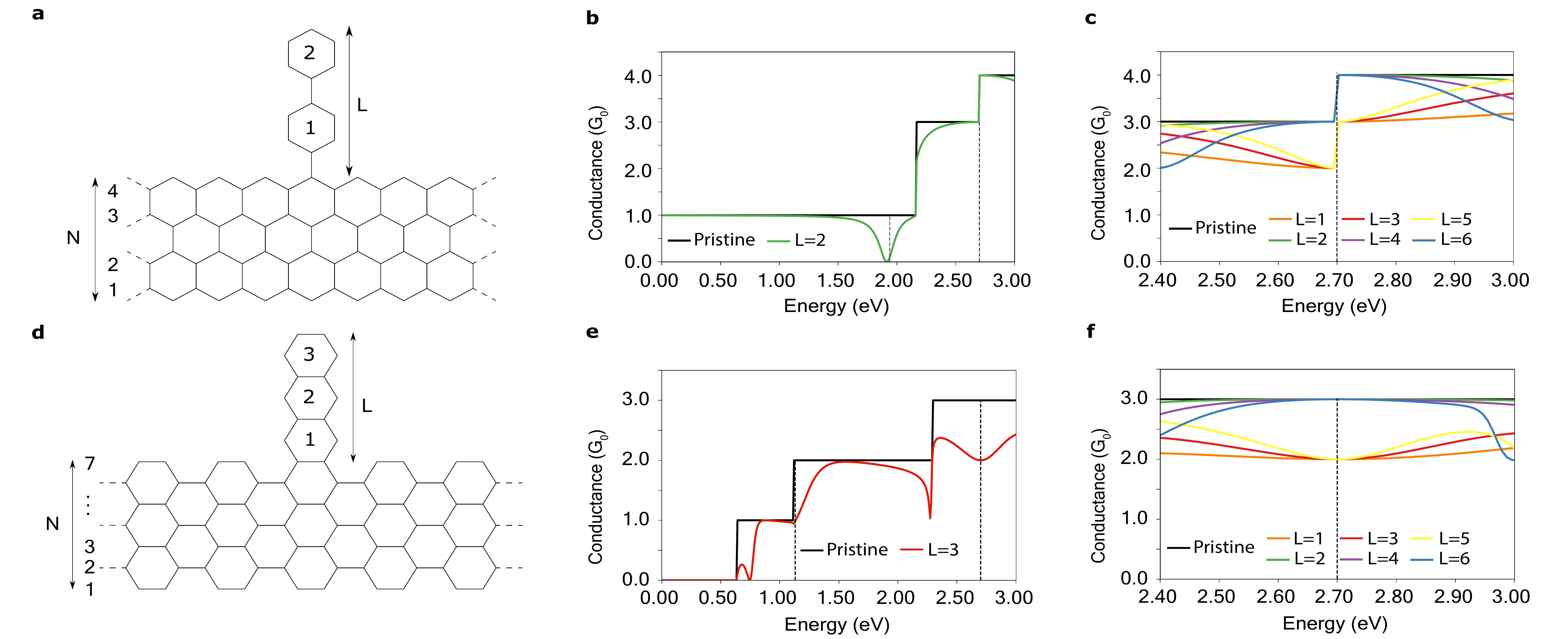}%
	\caption{(a) Atomic structure of 4-ZGNR  edge-functionalized with a $p$-polyphenylene group with $L=2$ (biphenyl). (b) Conductance spectra of 4-ZGNR hosting biphenyl group (L=2) with dashed lines indicating the electronic states of the molecule. (c) Conductance spectra of 4-ZGNR hosting $p$-polyphenylene groups with $1 \leq L \leq 6$ in the region of 2.70 eV.(d) Atomic structure of 7-AGNR edge-functionalized with a polyacene  group with $L=3$ (anthracene). (e) Conductance spectra of 7-AGNR hosting anthracene molecule (L=3) with dashed lines indicating the electronic states of the molecule. (f) Conductance spectra of 7-AGNR hosting polyacene groups with $1 \leq L \leq 6$ in the region of 2.70 eV.
		\label{fig1}}
\end{figure*} 

We start our investigation by relying on the 1NN-TB model Hamiltonians. In Figs.\ \ref{fig1}(a), (b) and (c), we show 4-ZGNR hosting a $p$-polyphenyl functional group of increasing lengths, ranging from phenyl (possessing a number of rings $L = 1$) to $p$-hexaphenyl ($L = 6$), along with its effect on the resulting conductance spectra. Only the conduction states are given, as the electron-hole symmetry is preserved in the 1NN-TB model in the entire energy spectrum. In Fig.\ \ref{fig1} (b) one clearly notices the presence of drop in the conductance at $E=1.94$ eV corresponding to the electronic state of the molecule, as well as a superimposition at $E=2.70$ eV to the conductance profile of that of the pristine system. Hence, the electronic state of the molecule at $E=2.70$ eV has no effect on the conductance profile in this particular case. We observe that depending on the number of rings in the guest molecule, distinct effects in the conductance spectra of the hosting nanoribbon emerges [see Fig.\ \ref{fig1}(c)]. We stress again that all $p$-polyphenyls host an electronic state at $E=2.70$ eV, but 4-ZGNR functionalized with $p$-polyphenyl groups possessing an even number of aromatic rings clearly shows conductance spectra approaching that of the the pristine nanoribbon at $E=2.70$ eV. On the other hand, the spectra upon the introduction of a $p$-polyphenyls with an odd number of rings display a Fano anti-resonance at $E=2.70$ eV. We unambiguously observe the even-odd effect in the vicinity ($\pm \delta E$) of the conductance step, and in the following we will present our results at $E=2.70 \pm \delta E$ (with $\delta E =0.001$ eV) to ensure avoidance of the irregularity. A  parallel conclusion can be drawn for 7-AGNR edge-functionalized with polyacenes displayed in Figs.\ \ref{fig1}(d), (e) and (f). In this case, however, the lack of a conductance step at $E=2.70$ eV makes Fano anti-resonances even more visible, as we show in Fig.\ \ref{fig1}(f). Again, resonant transport is observed when the edge-attached molecule possesses an even $L$, while an anti-resonance is seen for an odd $L$, which becomes narrower as the length of the chain increases, suggesting a weaker coupling to the nanoribbon \cite{Miro10}. Interestingly, the Fano anti-resonance happens at the position of a flat band in AGNRs as calculated by 1NN-TB model, whereas $E=2.70$ eV is the energy, where multiple sub-bands cross at $X$ point of the Brillouin zone in ZGNRs.

\begin{figure*}[t]
	\includegraphics[width=1.95\columnwidth]{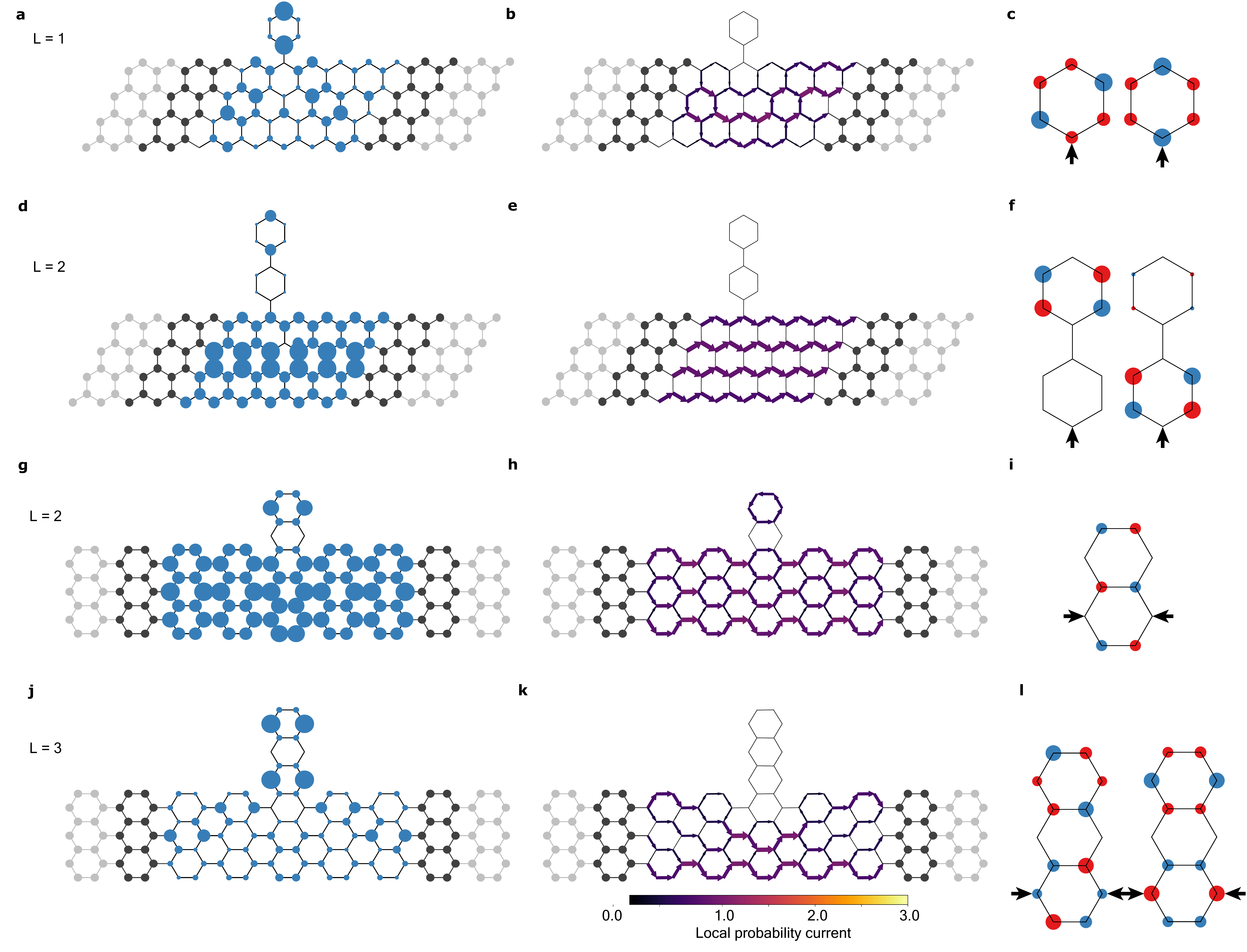}%
	\caption{(a) LDOS and (b) probability current at $E=2.70-\delta E$ eV (with $\delta E = 0.001$ eV) in 4-ZGNR edge-functionalized with a phenyl ($L=1$)  group. (c) Wave functions of the isolated  benzene  ($L=1$) molecule at $E=2.70$ eV, with colors indicating a phase difference of $\pi$ and black arrows indicating the binding sites.
		(d) LDOS and (e) probability current at $E=2.70-\delta E$ eV in 4-ZGNR edge-functionalized with biphenyl ($L=2$) group. (f) Wave functions of the isolated biphenyl molecule at $E=2.70$ eV. (g) LDOS and (h) probability current at $E=2.70$ eV in 7-AGNR edge-functionalized with naphtalene ($L=2$). (i) Wave functions of the isolated naphtalene molecule at $E=2.70$ eV. (j) LDOS and (k) probability current at $E=2.70$ eV in 7-AGNR edge-functionalized with anthracene ($L=3$). (l) Wave functions of the isolated anthracene molecule $E=2.70$ eV.
		\label{fig2}}
\end{figure*}

We complement our analysis of this fascinating even-odd phenomenon by presenting in Fig.\ \ref{fig2} the local densities of states (LDOS), local probability current maps, and the wave functions of 4-ZGNR  and 7-AGNR upon binding of $p$--polyphenyls ($L = 1, 2$) and polyacenes ($L=2, 3$), respectively, at the relevant energy $E=2.70$ eV.  The LDOS of 4-ZGNR functionalized with a phenyl group at the edge ($L=1$) shown in Fig.\ \ref{fig2}(a) indicates a pronounced localization at both the inner region of 4-ZGNR as well as at the phenyl group, whereas no density is observed on the edge site of ZGNR  which binds to the phenyl group. At this latter site, we also note the flow of the current is suppressed, see Fig.\ \ref{fig2}(b). Moving from phenyl to biphenyl ($L=2$), on the other hand, the LDOS is delocalized over the whole 4-ZGNR [Fig.\ \ref{fig2}(d)] and no density resides at the site which bridges the functional group to 4-ZGNR. Furthermore, the probability current remains unperturbed and features a symmetric flow through both edges of the nanoribbon, see Fig.\ \ref{fig2}(e). These results lend further support to the observations discussed in the previous paragraph, \emph{i.e.} that changes in the transport properties of edge-functionalized GNRs depend on the number of rings in functional group, with odd number of rings causing the disruption of the electronic structure at $E=2.70$ eV. Similar observations translate to 7-AGNR. In the case of naphtalene ($L=2$) [Fig.\ \ref{fig2}(g)], the LDOS is fully delocalized and no density can be found on the two atoms which bind the polyacene with the nanoribbon, hence displaying local currents resembling those of pristine 7-AGNR [Fig.\ \ref{fig2}(h)]. In contrast, functionalization with anthracene ($L = 3$) gives rise to strong localization which in turn hinders the current to flow between the two edge atoms in the nanoribbon at which the guest molecule binds, as seen in Figs.\ \ref{fig2}(j) and (k).

\begin{figure*}[t]
	\includegraphics[width=2\columnwidth]{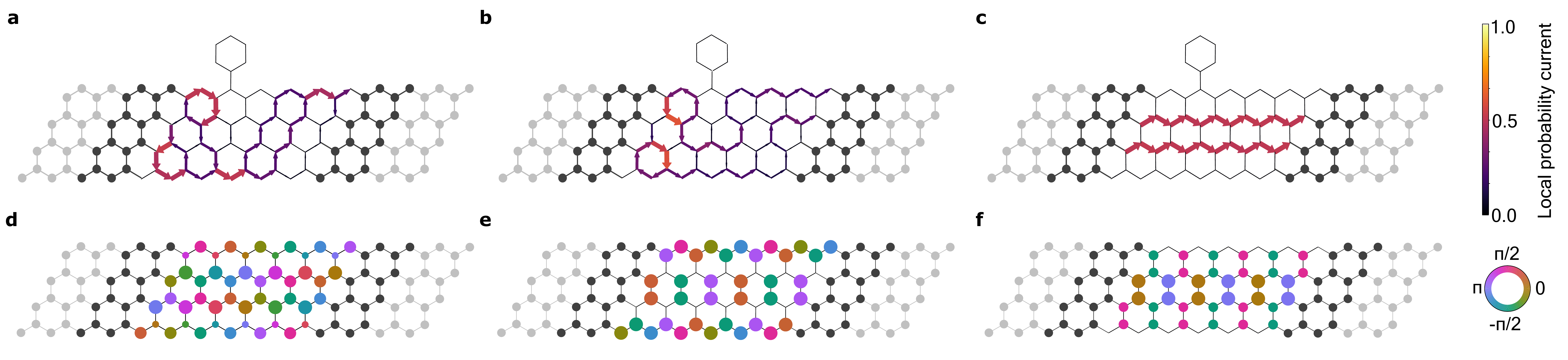}%
	\caption{Map of probability current of 4-ZGNR edge-functionalized with a phenyl group at $E=2.70$ eV for incoming (a) mode I, (b) mode II, and (c) mode III from the left lead. Wave function of 4-ZGNR at $E=2.70$ eV for incoming (d) mode I, (e) mode II, and (f) mode III from the left lead.}
	\label{fig3}
\end{figure*}

In the case of guest aromatic molecules with $L=2$, we observe that the state that is localized on the molecule is not interacting with the continuum of states of the nanoribbon, and the conductance at $E=2.70$ eV is the same as that of the ideal lead [Figs.\ \ref{fig2}(d),(g)]. The reason for this traces back to the spatial distribution of the electronic states in the corresponding isolated molecules at $E=\pm 2.70$ eV, as given in Figs.\ \ref{fig2}(c), (f), (i), and (j). At variance with benzene [Fig.\ \ref{fig2}(c)]], in the biphenyl group the wave function [Fig.\ \ref{fig2}(f)] does not localize on the atoms that are directly bound to GNRs, thereby preventing any interactions with the continuum of states of GNR to occur. In general, the spatial distribution of the wave function is similar in all $p$-polyphenyl molecules possessing an even number of rings, showing zero-weight at the binding sites and strong localization along the molecular armchair edges. On the other hand, aromatic molecules featuring an odd number of rings feature a wave function that does localize on the sites binding to the nanoribbon, thus leading to interaction with the ZGNR states and resulting in Fano anti-resonances located at $E=\pm 2.70$ eV. Differently from $p$-polyphenyls, in the case of polyacenes groups shown in Figs.\ \ref{fig2}(i) and (j), there exist two sites which bind to 7-AGNR. Similarly to $p$-polyphenyls, however, depending on whether the wave function does reside or not on the binding sites, either an intact or disrupted transport emerges in the current maps. Hence, we conclude that the localization of the wave function at the sites of the guest functional groups which bind to the GNRs is the key ingredient in governing the destructive interference of the electron transport in the hosting nanoribbon.

As shown in Figs. \ref{fig1}((c) and (f), edge-functionalization of the nanoribbon with a molecule containing an odd number of aromatic rings removes exactly one quantum from the conductance at $E=\pm 2.70$ eV. Furthermore, we see that the local currents in Figs.\ \ref{fig2}(b) and (k) break the symmetry along the edges and hence indicate a different response from the incoming modes. For the representative case of 4-ZGNR edge-functionalized with a phenyl group ($L=1$), we present in Figs.\ \ref{fig3} (a), (b), and (c) the current maps due to the three incoming modes from the left lead at $E=2.70-\delta E$. Our results demonstrate that, while the current is disrupted in panels (a) and (b), it is fully preserved in panel (c). The origin of this effect can be rationalized by inspecting the behavior of the incoming wave function, as shown in Figs.\ \ref{fig3}(d),(e), and (f). Of the three wave functions displayed, only the one presented in Fig.\ \ref{fig3}(f) does not exhibit finite weight on the atoms at the edge of the nanoribbon, thereby indicating that the local current in the inner region of the nanoribbon is not affected by the guest molecule.  On the other hand,  the wave functions shown in \ref{fig3}(d) and (e) are localized to some degree on the edge atoms as well.  This is further reflected  in the current maps of Fig. \ref{fig3}(a) and (b), in which one can observe how the local current is not flowing through the site that is bound to the phenyl group. A similar behavior has been previously reported in carbon-based $sp^2$-hybridized structures whereby one atom experiences a large on-site potential \cite{Chen18a}. We stress that the same effect is displayed by 7-AGNR edge-functionalized with a group containing an odd number of rings (not shown here), with the difference that in this latter case the guest molecule is covalently bound through two distinct carbon sites as compared to 4-ZGNR. Remarkably, we demonstrate that the even-odd phenomenon boils down to the interaction between two (four) sites, \emph{i.e.}\ one (two) on the ZGNR (AGNR) and one (two) on the edge-attached aromatic molecule.

\begin{figure*}[t]
	\includegraphics[width=2\columnwidth]{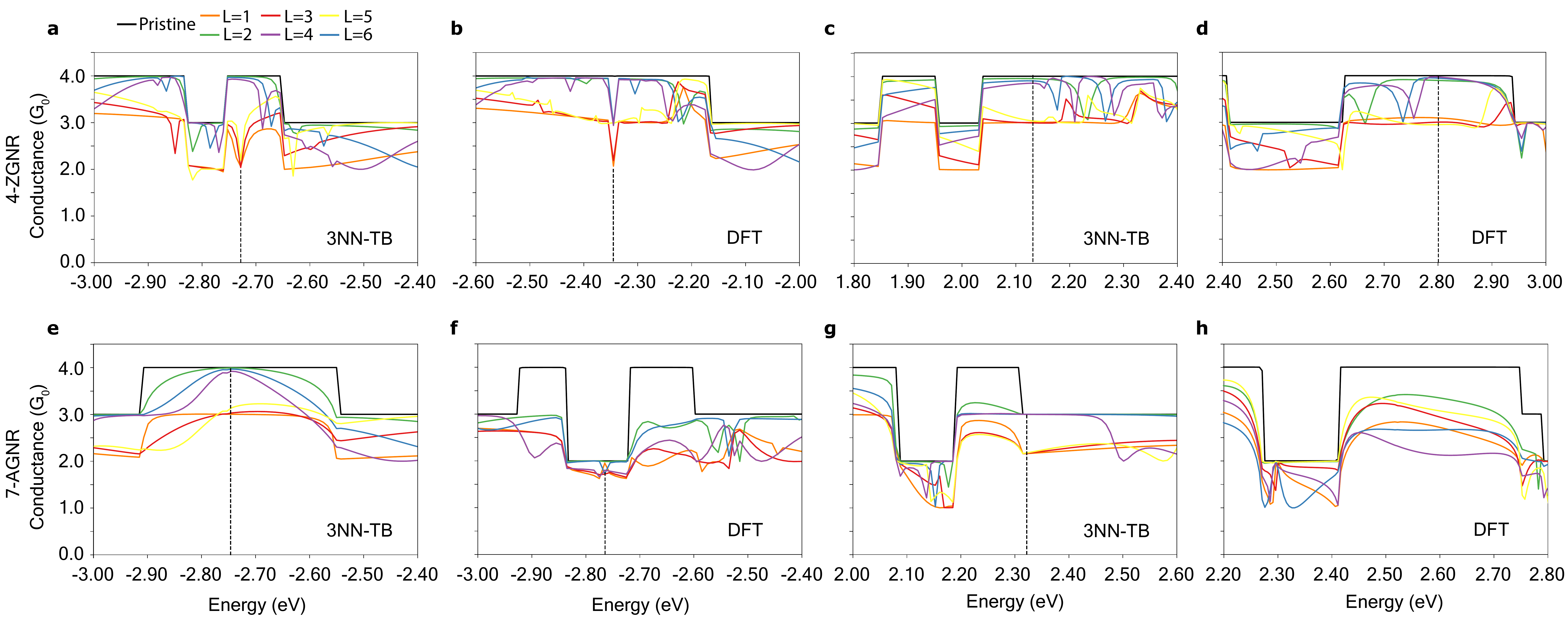}%
	\caption{Conductance spectra of  4-ZGNR edge-functionalized with $p$-polyphenylene groups of $1 \leq L \leq 6$ below the Fermi level at the (a)  3NN-TB,  and (b) DFT levels, as well as above the Fermi level at the (c) 3NN-TB and (d)  DFT levels. Conductance spectra of 7-AGNR edge-functionalized with polyacene groups of $1 \leq L \leq 6$ below the Fermi level at the (e)  3NN-TB,  and (f) DFT levels, as well as above the Fermi level at the (g) 3NN-TB and (h) DFT levels.}
	\label{fig4}
\end{figure*}

Next, we expand our investigation by employing more realistic models, namely third nearest-neighbor (3NN) TB model and DFT calculations in order to verify that the observed even-odd effect is not an artifact of the possibly oversimplified 1NN-TB model. In Fig.\ \ref{fig4}, we present the conductance spectra of 4-ZGNR and 7-AGNR hosting $p$-polyphenyl and polyacene groups of length $ 1 \leq L \leq 6$  bound to the edge. As expected, 3NN-TB model and DFT calculations break the electron-hole symmetry observed in the 1NN-TB model. Hence, both valence and conduction states are presented. In Figs. \ref{fig4}(a) and (c), we observe that the even-odd effect is retained when the 3NN-TB model is adopted for 4-ZGNR system both in the valence and conduction states, even though the energy at which it takes place is shifted by $\sim$0.6 eV with respect to the 1NN-TB model for the latter. Additionally, we remark that there are no distinct resonances and anti-resonances occurring at a particular energy, rather such effect appears within the energy window in which the conductance takes approximately the value of 4.0 or 3.0 G$_{0}$. Furthermore, DFT results shown in Figs. \ref{fig4}(b) and (d) also allow one to discriminate between odd and even values of $L$, both in valence and conduction states, and, similar to the 3NN-TB model,  the phenomenon is again slightly shifted in energy w.r.t.\ the 1NN-TB model, yet to a different extent as compared the 3NN-TB results. The conductance spectra of edge-functionalized 7-AGNR obtained at the 3NN-TB and DFT levels are given in Figs.\ \ref{fig4}(e) and (f), respectively. While the even-odd effect is observed in the 3NN-TB model calculations, we found that, upon increasing $L$, the conductance deviates from the the resonance and anti-resonance behaviors observed in the 1NN-TB model. Hence, the even-odd effect becomes difficult to be resolved in longer aromatic functional groups. Furthermore, DFT calculations of the valence states yield only a weak separation between odd and even $L$  at $E \approx -2.80$ eV, though for even values of  $L$  resonant transport is preserved. The 3NN-TB results for the conduction states shown in Fig.\ \ref{fig4}(g) reveal that the effect can still be clearly observed, but, similar to 4-ZGNR, it appears shifted to $E=2.30$ eV and spread over a wider energy window. Finally, we did not single out any even-odd effect characteristics for 7-AGNR with side-attached aromatic molecules in the conduction states calculated by means of DFT. Overall, we establish that the even-odd phenomenon is not an artifact of the 1NN-TB model, though it appears largely exaggerated by this simplified approach.

\begin{figure*}[t]
	\includegraphics[width=2\columnwidth]{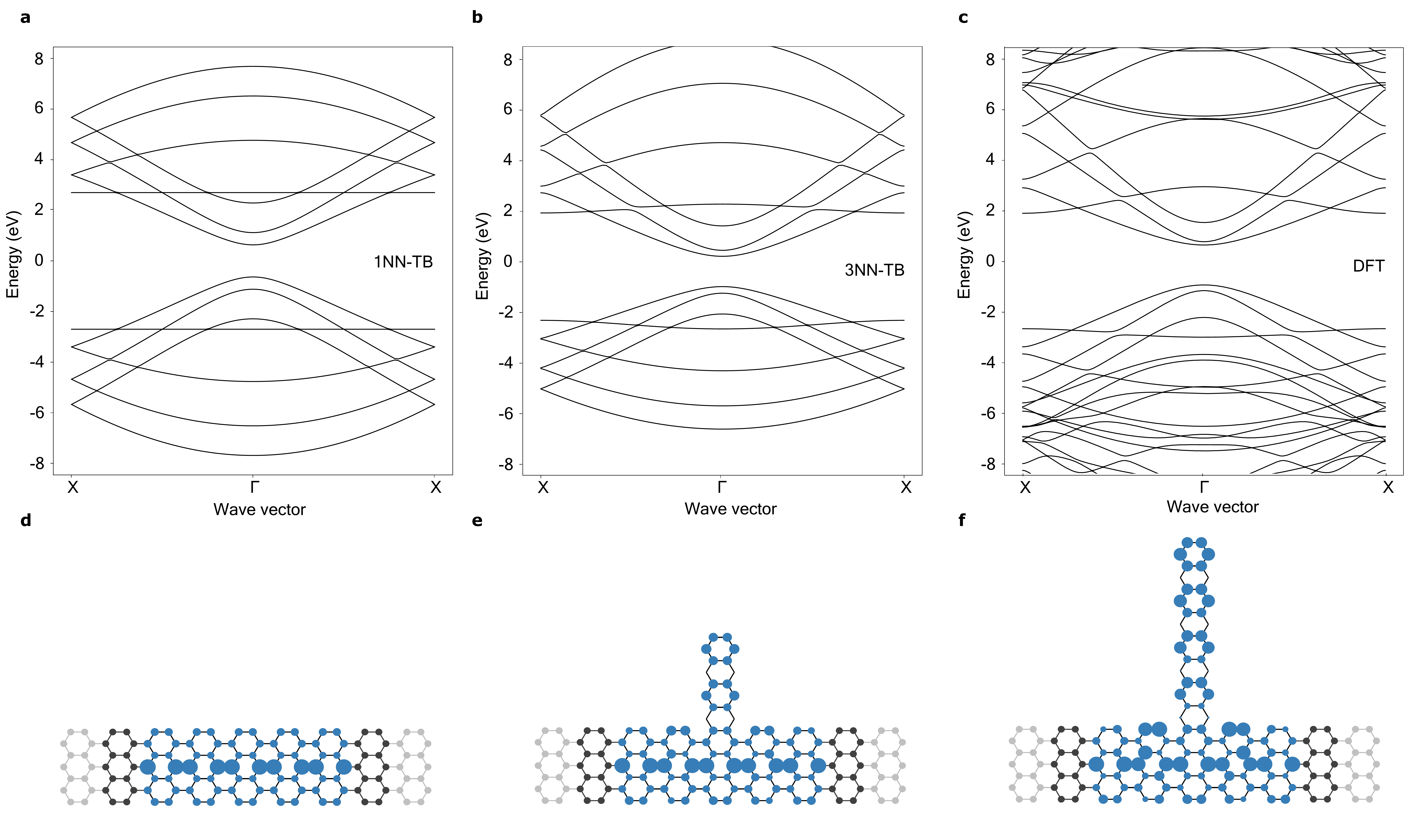}%
	\caption{Electronic band structure of pristine 7-AGNR obtained at the (a) 1NN-TB, (b) 3NN-TB, and (c) DFT levels. (d) 3NN-TB model LDOS at $E=-2.70$ eV for (d) pristine, (e) edge-functionalized with tetracene ($L=4$), and  (f) edge-functionalized with octacene ($L=8$) 7-AGNR.}
	\label{fig5}
\end{figure*}

We then determine the band structure of pristine 7-AGNR to rationalize the discrepancy in the even-odd effect emerging in the different adopted levels of theory. Our results are presented in Fig.\ \ref{fig5}. The band structure obtained at the 1NN-TB level exhibits a flat band at $E=\pm 2.70$ eV [see Fig.\ \ref{fig5}(a)], \emph{i.e.}\ where the even-odd effect  occurs. On the other hand, both the 3NN-TB model and DFT calculations reveal that such band acquires some dispersion character and slightly shifts in energy as compared to the 1NN-TB result.  Such a poor description of the band structure of  7-AGNR at the 1NN-TB level explains why the even-odd effect is shifted away from $E=2.70$ eV in higher levels of theory, and also accounts for the fact that this phenomenon is observed at wider energy windows which correspond to the bandwidth.  Furthermore, at variance with the 1NN-TB model, the DFT band structure [Fig.\ \ref{fig5}(c)] clearly shows strongly dispersed character of the conduction states, hence substantially quenching the even-odd effect in this energy range.

Finally, we discuss the observed deviation from the conductance values of 4.0 and 3.0 G$_{0}$ shown in Fig.\ \ref{fig4}(e), for even and odd values of $L$, respectively, on the basis of the 3NN-TB model LDOS of 7-AGNR systems at $E = 2.70$ eV. We compare the LDOS of the pristine system [Fig.\ \ref{fig5}(d)] with that of the nanoribbon hosting a tetracene ($L = 4$) [Fig.\ \ref{fig5}(e)] or an octacene ($L = 8$) [Fig.\ \ref{fig5}(f)]. Our findings indicate that the LDOS in the inner region of the 7-AGNR edge-functionalized with tetracene closely resembles that of the pristine system, with the LDOS on the functional group being well separated from the 7-AGNR and resulting in resonant transport. The increase in the number of the rings, \emph{e.g} moving from tetracene to octacene, is accompanied by a disruption in the LDOS in the lead, as shown in Fig.\ \ref{fig5}(f). In addition, the LDOS starts to develop on the two binding sites, hence detrimentally perturbing the transport properties and suppressing its resonant character. We conclude that a further increase in the length of acenes containing an even number of rings promotes a significant hybridization of the localized state on the guest molecule with the continuum of states in the hosting AGNR, yielding a decrease in the conductance.

\section{Summary and Conclusions}
\label{conclusions} 

In summary, we have carried out a theoretical investigation of the transport properties of GNRs with edges chemically functionalized with $p$-polyphenyl and polyacene groups. Our 1NN-TB results indicate that, depending on whether the number of aromatic rings in the guest molecule is even or odd, either constructive or destructive interference takes place at $E=\pm t_1=\pm 2.70$ eV in the conductance spectrum. In the case of functional groups with an even number of rings, this effect stems from the fact that the local density of states does not reside on the sites of the guest molecule which covalently bind to the nanoribbon, resulting in negligible interactions between the continuum of states in GNR and the localized state on the edge-attached molecule. On the other hand, upon binding of functional groups containing an odd number of rings to the nanoribbon, the electron transport is decreased by one quantum of conductance at $E=t_1$. Further analysis indicates that such functionalization affects local current and hence conductance to an extent which is governed by the behavior of the wave function in the lead.

We then established that the even-odd phenomenon is largely preserved when adopting higher levels of theory, \emph{i.e.}\  3NN-TB model and DFT. As compared to the simplified 1NN-TB model, however, this phenomenon is shifted in energy and spread over a larger energy window, as a consequence of the very approximate nature of the band structure obtained at the 1NN-TB level.  Also, we have suggested that such even-odd effect becomes less pronounced as the number of aromatic rings in the guest molecule increases.

In summary, we have revisited and clarified the origin of the even-odd conductance effect observed in graphene nanoribbons with armchair or zigzag edges chemically functionalized with aromatic functional groups. We have provided a detailed understanding of the interplay between the localized states on the guest molecules and the continuum of states of the hosting graphene nanoribbon, and on its role in governing the resulting electron transport. Overall, our results promote the validity of graphene nanoribbons as promising candidates for chemosensing devices, and offer a theoretical insight into the formulation of guidelines towards their realization.

\section{Acknowledgments}
The authors are financially supported by the Swiss National Science Foundation (Grant No.\ 172534) and the NCCR MARVEL. First-principles calculations have been performed at the Swiss National Supercomputing Center (CSCS) under the projects s832 and s1008.

\bibliography{GNR_bib}

 \end{document}